\documentstyle[amssymb,11pt,epsfig,psfig,amsmath]{article}

\begin{document}

\title{Casimir densities for a spherical brane in Rindler-like spacetimes}
\author{ A. A. Saharian$^{1}$\thanks{%
E-mail: saharyan@server.physdep.r.am } and M. R. Setare$^{2}$\thanks{%
E-mail: rezakord@ipm.ir} \\
{\it $^1$ Department of Physics, Yerevan State University, Yerevan, Armenia }
\\
{\it $^2$ Institute for Theoretical Physics and Mathematics, Tehran, Iran}}
\maketitle

\begin{abstract}
Wightman function, the vacuum expectation values of the field square and the
energy-momentum tensor are evaluated for a scalar field obeying mixed
boundary condition on a spherical brane in $(D+1)$-dimensional Rindler-like
spacetime $Ri\times S^{D-1}$, where $Ri$ is a two-dimensional Rindler
spacetime. This spacetime approximates the near horizon geometry of $(D+1)$%
-dimensional black hole in the large mass limit. The vacuum expectation
values are presented as the sum of boundary-free and brane-induced parts.
Further we extract from the Wightman function for the boundary-free geometry
the corresponding function in the bulk $R^{2}\times S^{D-1}$. For the latter
geometry the vacuum expectation values of the field square and the
energy-momentum tensor do not depend on the spacetime point. For the
renormalization of these quantities we use zeta regularization technique.
Various limiting cases of the brane-induced vacuum expectation values are
investigated.
\end{abstract}

\bigskip

PACS number(s): 03.70.+k, 04.62.+v, 11.10.Kk

\section{Introduction}

\label{sec:Int}

Motivated by string/M theory, the AdS/CFT correspondence, and the
hierarchy problem of particle physics, braneworld models were
studied actively in recent years \cite{Hora96}-\cite{Rand99}. In
this models, our universe is realized as a boundary of a higher
dimensional spacetime. In particular, a well studied example is when
the bulk is an AdS space. In the cosmological context, embedding of
a four dimensional Friedmann-Robertson-Walker universe was also
considered when the bulk is described by AdS or AdS black hole \cite%
{Nihe99,AdSbhworld}. In the latter case, the mass of the black hole
was found to effectively act as an energy density on the brane with
the same equation of state of radiation. Representing radiation as
conformal matter and exploiting AdS/CFT correspondence, the
Cardy-Verlinde formula \cite{Verl00} for the entropy was found for
the universe (for the entropy formula in the case of dS black hole
see \cite{Seta02}). Moreover, in the AdS/CFT correspondence, the
case of a bulk AdS black hole represents a different phase of the
same theory and there is the exciting connection that a transition
between an ordinary bulk AdS and a bulk AdS black hole corresponds
to the confinement-de confinement transition in the dual CFT
\cite{Witt98}. Therefore it seems interesting to generalize the
study of quantum effects due to bulk AdS black holes.

The investigation of quantum effects in braneworld models is of considerable
phenomenological interest, both in particle physics and in cosmology. The
braneworld corresponds to a manifold with dynamical boundaries and all
fields which propagate in the bulk will give Casimir-type contributions to
the vacuum energy (for reviews of the Casimir effect see Refs. \cite{Most97}%
), and as a result to the vacuum forces acting on the branes. In
dependence of the type of a field and boundary conditions imposed,
these forces can either stabilize or destabilize the braneworld. In
addition, the Casimir energy gives a contribution to both the brane
and bulk cosmological constants and, hence, has to be taken into
account in the self-consistent formulation of the braneworld
dynamics. Motivated by these, the role of quantum effects in
braneworld scenarios has received a great deal of attention. For a
conformally coupled scalar this effect was initially studied in Ref.
\cite{Fabi00} in the context of M-theory, and subsequently in Refs.
\cite{Noji00a,Flac01} for a background Randall--Sundrum geometry.
The models with dS and AdS branes, and higher dimensional brane
models are considered as well \cite{Noji00b}.

In view of these recent developments, it seems interesting to
generalize the study of quantum effects to other types of bulk
spacetimes. In particular, it is of interest to consider
non-Poincar\'{e} invariant braneworlds, both to better understand
the mechanism of localized gravity and for possible cosmological
applications. Bulk geometries generated by higher-dimensional black
holes are of special interest. In these models , the tension and the
position of the brane are tuned in terms of black hole mass and
cosmological constant and brane gravity trapping occurs in just the
same way as in the Randall-Sundrum model. Braneworlds in the
background of the AdS black hole were studied in \cite{AdSbhworld}.
Like pure AdS space the AdS black hole may be superstring vacuum. It
is of interest to note that the phase transitions which may be
interpreted as confinement-deconfinement transition in AdS/CFT setup
may occur between pure AdS and AdS black hole \cite{Witt98}. Though,
in the generic black hole background the investigation of
brane-induced quantum effects is techniqally complicated, the exact
analytical results can be obtained in the near horizon and large
mass limit when the brane is close to the black hole horizon. In
this limit the black hole geometry may be approximated by the
Rindler-like manifold (for some investigations of quantum effects on
background of Rindler-like spacetimes see \cite{Byts96} and
references therein). In the present paper we investigate the
Wightman function, the vacuum expectation values of the field square
and the energy-momentum tensor for a scalar field with an
arbitrary curvature coupling parameter for the spherical brane on the bulk $%
Ri\times S^{D-1}$, where $Ri$ is a two-dimensional Rindler
spacetime. Note that the corresponding quantities induced by a
single and two parallel flat branes in the bulk geometry $Ri\times
R^{D-1}$ for both scalar and electromagnetic fields are investigated
in \cite{Cand77}. This problem is also of separate interest as an
example with gravitational and boundary-induced polarizations of the
vacuum, where all calculations can be performed in a closed form.
The paper is organized as follows. In section \ref{sec:WF} we
consider the positive frequency Wightman function in the region
between the brane and Rindler horizon. This function is presented as
the sum of boundary-free and boundary-induced parts. The vacuum
expectation values for the boundary-free geometry are investigated
in section \ref{sec:bfree}. The vacuum expectation values induced by
a spherical brane are studied in section \ref{sec:VEVEMT}. Section
\ref{sec:Conc} summarizes the main results of the paper.

\section{Wightman function}

\label{sec:WF}

Let us consider a scalar field $\varphi (x)$ propagating on background of $%
(D+1)$-dimensional Rindler-like spacetime $Ri\times S^{D-1}$, where $Ri$ is
a two-dimensional Rindler spacetime. The corresponding metric is described
by the line element%
\begin{equation}
ds^{2}=\xi ^{2}d\tau ^{2}-d\xi ^{2}-r_{H}^{2}d\Sigma _{D-1}^{2},
\label{ds22}
\end{equation}%
with the Rindler-like $(\tau ,\xi )$ part and $d\Sigma _{D-1}^{2}$
is the line element for the space with positive constant curvature
with the Ricci scalar $R=(D-2)(D-1)/r_{H}^{2}$. Line element
(\ref{ds22}) describes the near horizon geometry of
$(D+1)$-dimensional topological black hole with the line
element \cite{Mann97}%
\begin{equation}
ds^{2}=A_{H}(r)dt^{2}-\frac{dr^{2}}{A_{H}(r)}-r^{2}d\Sigma _{D-1}^{2},
\label{ds21}
\end{equation}%
where%
\begin{equation}
A_{H}(r)=k+\frac{r^{2}}{l^{2}}-\frac{r_{0}^{n+2}}{l^{2}r^{n}},\quad n=D-2,
\label{Ar}
\end{equation}%
and the parameter $k$ classifies the horizon topology, with $k=0,-1,1$
corresponding to flat, hyperbolic, and elliptic horizons, respectively. In (\ref%
{Ar}) the parameter $l$ is related to the bulk cosmological constant and the
parameter $r_{0}$ depends on the mass of the black hole and on the bulk
gravitational constant. In the non extremal case the function $A_{H}(r)$ has
a simple zero at $r=r_{H}$. In the near horizon limit, introducing new
coordinates $\tau $ and $\rho $ in accordance with%
\begin{equation}
\tau =\frac{1}{2}A_{H}^{\prime }(r_{H})t,\quad r-r_{H}=\frac{1}{4}%
A_{H}^{\prime }(r_{H})\xi ^{2},  \label{tau}
\end{equation}%
the line element is written in the form (\ref{ds22}). Note that for a $(D+1)$%
-dimensional Schwarzschild black hole \cite{Call88} one has $%
A_{H}(r)=1-(r_{H}/r)^{D-2}$ and, hence, $A_{H}^{\prime }(r_{H})=n/r_{H}$.

The field equation is in the form%
\begin{equation}
\left( g^{ik}\nabla _{i}\nabla _{k}+m^{2}+\zeta R\right) \varphi (x)=0,
\label{fieldeq1}
\end{equation}%
where $\zeta $ is the curvature coupling parameter. Below we will assume
that the field satisfies the Robin boundary condition on the hypersurface $%
\xi =a$:
\begin{equation}
\left( A+B\frac{\partial }{\partial \xi }\right) \varphi =0,\quad \xi =a,
\label{bound1}
\end{equation}%
with constant coefficients $A$ and$\ B$. The Dirichlet and Neumann
boundary conditions are obtained as special cases. In accordance
with \ (\ref{tau}), the hypersurface $\xi =a$ corresponds to the
spherical shell near the black hole horizon with the radius
$r_{a}=r_{H}+A_{H}^{\prime }(r_{H})a^{2}/4$. Here we consider the
general case of the ratio $A/B$. The application to the braneworld
scenario will be given below.

To evaluate the vacuum expectation values of the field square and
the energy-momentum tensor we need a complete set of eigenfunctions
satisfying the boundary condition (\ref{bound1}). Below we shall use
the hyperspherical angular coordinates $(\vartheta ,\phi )=(\theta
_{1},\theta _{2},\ldots
,\theta _{n},\phi )$ on $S^{D-1}$\ with $0\leq \theta _{k}\leq \pi $, $%
k=1,\ldots ,n$, and $0\leq \phi \leq 2\pi $. In these coordinates the
variables are separated and the eigenfunctions can be written in the form%
\begin{equation}
\varphi _{\alpha }(x)=C_{\alpha }f(\xi )Y(m_{k};\vartheta ,\phi )e^{-i\omega
\tau },  \label{eigfunc1}
\end{equation}%
where $m_{k}=(m_{0}\equiv l,m_{1},\ldots m_{n})$, and $m_{1},m_{2},\ldots
m_{n}$ are integers such that%
\begin{equation}
0\leq m_{n-1}\leq \cdots \leq m_{1}\leq l,\quad -m_{n-1}\leq m_{n}\leq
m_{n-1},  \label{mk}
\end{equation}%
$Y(m_{k};\vartheta ,\phi )$ is the surface harmonic of degree $l$ \cite%
{Erdelyi}. Substituting this into Eq. (\ref{fieldeq1}) we see that the
function $f(\xi )$ satisfies the equation%
\begin{equation}
\xi \frac{d}{d\xi }\left( \xi \frac{df}{d\xi }\right) +\left( \omega
^{2}-\xi ^{2}\lambda _{l}^{2}\right) f(\xi )=0,  \label{feq}
\end{equation}%
with the notation%
\begin{equation}
\quad \lambda _{l}=\frac{1}{r_{H}}\sqrt{l(l+n)+\zeta n(n+1)+m^{2}r_{H}^{2}}.
\label{lambdal}
\end{equation}%
The linearly independent solutions to (\ref{feq}) are the Bessel modified
functions $I_{\pm i\omega }(\lambda _{l}\xi )$ and $K_{i\omega }(\lambda
_{l}\xi )$ with the imaginary order.

In the region $0<\xi <a$ the solution to (\ref{feq}) satisfying boundary
condition (\ref{bound1}) has the form%
\begin{equation}
f(\xi )=Z_{i\omega }(\lambda _{l}\xi ,\lambda _{l}a)\equiv K_{i\omega
}(\lambda _{l}\xi )-\frac{\bar{K}_{i\omega }(\lambda _{l}a)}{\bar{I}%
_{i\omega }(\lambda _{l}a)}I_{i\omega }(\lambda _{l}\xi ),  \label{f2}
\end{equation}%
where for a given function $F(z)$ we use the notation%
\begin{equation}
\bar{F}(z)=AF(z)+bzF^{\prime }(z)=0,\quad b=B/a.  \label{fbarnot}
\end{equation}%
The coefficient $C_{\alpha }$ in (\ref{eigfunc1}) can be found from the
normalization condition%
\begin{equation}
\int \left\vert \varphi _{\alpha }(x)\right\vert ^{2}\sqrt{-g}dV=\frac{1}{%
2\omega },  \label{normcond}
\end{equation}%
where the integration goes over the region between the horizon and the
sphere. Substituting eigenfunctions (\ref{eigfunc1}), using the relation
\begin{equation}
\int \left\vert Y(m_{k};\vartheta ,\phi )\right\vert ^{2}d\Omega =N(m_{k})
\label{normsph}
\end{equation}%
for spherical harmonics, one finds%
\begin{equation}
C_{\alpha }=\frac{1}{\pi }\sqrt{\frac{\sinh \omega \pi }{r_{H}^{n+1}N(m_{k})}%
}.  \label{Calfa}
\end{equation}%
The explicit form for $N(m_{k})$ is given in \cite{Erdelyi} and will not be
necessary for the following considerations in this paper.

First of all we evaluate the positive frequency Wightman function%
\begin{equation}
G^{+}(x,x^{\prime })=\langle 0\left\vert \varphi (x)\varphi (x^{\prime
})\right\vert 0\rangle ,  \label{W1}
\end{equation}%
where $|0\rangle $ is the amplitude for the corresponding vacuum state. By
expanding the field operator over eigenfunctions and using the commutation
relations one can see that%
\begin{equation}
G^{+}(x,x^{\prime })=\sum_{\alpha }\varphi _{\alpha }(x)\varphi _{\alpha
}^{\ast }(x^{\prime }).  \label{W2}
\end{equation}%
Substituting eigenfunctions (\ref{eigfunc1}) with the function (\ref{f2})
into this mode sum and by making use the addition theorem%
\begin{equation}
\sum_{m_{k}}\frac{1}{N(m_{k})}Y(m_{k};\vartheta ,\phi )Y(m_{k};\vartheta
^{\prime },\phi ^{\prime })=\frac{2l+n}{nS_{D}}C_{l}^{n/2}(\cos \theta ),
\label{addtheor}
\end{equation}%
for the Wightman function one finds%
\begin{eqnarray}
G^{+}(x,x^{\prime }) &=&\frac{r_{H}^{1-D}}{\pi ^{2}nS_{D}}\sum_{l=0}^{\infty
}(2l+n)C_{l}^{n/2}(\cos \theta )  \nonumber \\
&&\times \int_{0}^{\infty }d\omega \sinh (\omega \pi )e^{-i\omega (\tau
-\tau ^{\prime })}Z_{i\omega }(\lambda _{l}\xi ,\lambda _{l}a)Z_{i\omega
}^{\ast }(\lambda _{l}\xi ^{\prime },\lambda _{l}a).  \label{W3}
\end{eqnarray}%
In (\ref{addtheor}), $S_{D}=2\pi ^{D/2}/\Gamma (D/2)$ is the total area of
the surface of the unit sphere in $D$-dimensional space, $C_{l}^{n/2}(x)$ is
the Gegenbauer or ultraspherical polynomial of degree $l$ and order $n/2$, $%
\theta $ is the angle between directions $(\vartheta ,\phi )$ and $%
(\vartheta ^{\prime },\phi ^{\prime })$, and the sum is taken over the
integer values $m_{k},\,k=1,2\ldots $ in accordance with (\ref{mk}). To
transform the expression on the right of (\ref{W3}), we present the product
of the functions $Z_{i\omega }$ in the form%
\begin{eqnarray}
Z_{i\omega }(\lambda \xi ,\lambda a)Z_{i\omega }^{\ast }(\lambda \xi
^{\prime },\lambda a) &=&K_{i\omega }(\lambda \xi )K_{i\omega }(\lambda \xi
^{\prime })+\frac{\pi \bar{K}_{i\omega }(\lambda a)}{2i\sinh \pi \omega }
\nonumber \\
&&\times \sum_{\sigma =-1,1}\frac{I_{i\sigma \omega }(\lambda \xi
)I_{i\sigma \omega }(\lambda \xi ^{\prime })}{\sigma \bar{I}_{i\sigma \omega
}(\lambda a)}.  \label{form1}
\end{eqnarray}%
On the base of this formula from (\ref{W3}) one finds%
\begin{eqnarray}
G^{+}(x,x^{\prime }) &=&G_{0}^{+}(x,x^{\prime })+\frac{r_{H}^{1-D}}{2i\pi
nS_{D}}\sum_{l=0}^{\infty }(2l+n)C_{l}^{n/2}(\cos \theta )  \nonumber \\
&&\times \int_{0}^{\infty }d\omega e^{-i\omega (\tau -\tau ^{\prime })}\bar{K%
}_{i\omega }(\lambda _{l}a)\sum_{\sigma =-1,1}\frac{I_{i\sigma \omega
}(\lambda _{l}\xi )I_{i\sigma \omega }(\lambda _{l}\xi ^{\prime })}{\sigma
\bar{I}_{i\sigma \omega }(\lambda _{l}a)},  \label{W4}
\end{eqnarray}%
where the part
\begin{eqnarray}
G_{0}^{+}(x,x^{\prime }) &=&\frac{r_{H}^{1-D}}{\pi ^{2}nS_{D}}%
\sum_{l=0}^{\infty }(2l+n)C_{l}^{n/2}(\cos \theta )  \nonumber \\
&&\times \int_{0}^{\infty }d\omega \sinh (\omega \pi )e^{-i\omega (\tau
-\tau ^{\prime })}K_{i\omega }(\lambda _{l}\xi )K_{i\omega }(\lambda _{l}\xi
^{\prime })  \label{W0}
\end{eqnarray}%
does not depend on the parameter $a$ determining the radius of the spherical
shell and corresponds to the Wightman function in the situation when the
spherical shell is absent. Assuming that the function $\bar{I}_{i\omega
}(\lambda a)$ ($\bar{I}_{-i\omega }(\lambda a)$) has no zeros for $-\pi
/2\leq {\rm arg}\,\omega <0$ ($0<{\rm arg}\,\omega <\pi /2$) we can rotate
the integration contour over $\omega $ by angle $-\pi /2$ for the term with $%
\sigma =1$ and by angle $\pi /2$ for the term with $\sigma =-1$. The
integrals taken around the arcs of large radius tend to zero under the
condition $|\xi \xi ^{\prime }|<a^{2}e^{|\tau -\tau ^{\prime }|}$ (note
that, in particular, this is the case in the coincidence limit for the
region under consideration). As a result for the Wightman function one
obtains%
\begin{equation}
G^{+}(x,x^{\prime })=G_{0}^{+}(x,x^{\prime })+\langle \varphi (x)\varphi
(x^{\prime })\rangle ^{(b)},  \label{W5}
\end{equation}%
where for the sphere-induced part one has%
\begin{eqnarray}
\langle \varphi (x)\varphi (x^{\prime })\rangle ^{(b)} &=&-\frac{r_{H}^{1-D}%
}{\pi nS_{D}}\sum_{l=0}^{\infty }(2l+n)C_{l}^{n/2}(\cos \theta )  \nonumber
\\
&&\times \int_{0}^{\infty }d\omega \frac{\bar{K}_{\omega }(\lambda _{l}a)}{%
\bar{I}_{\omega }(\lambda _{l}a)}I_{\omega }(\lambda _{l}\xi )I_{\omega
}(\lambda _{l}\xi ^{\prime })\cosh [\omega (\tau -\tau ^{\prime })].
\label{Wb1}
\end{eqnarray}%
For the points away the brane this part is finite in the coincidence limit.

We have investigated the Whightman function in the region between
the horizon and the boundary located at $\xi =a$ for an arbitrary
ratio of boundary coefficients $A/B$. In the corresponding
braneworld scenario the geometry is made up by two slices of the
region $0<\xi <a$ glued together at the brane with a orbifold-type
symmetry condition analogous to that in the Randall-Sundrum model
and the ratio $A/B$ for bulk scalars is related to the brane mass
parameter of the field and the extrinsic curvature of the brane. The
corresponding formula is obtained by the way similar to that in the
case of the Randall-Sundrum braneworld (see, for instance,
\cite{Flac01,Gher00}). For this we note that in braneworlds the
action for a scalar field with general curvature coupling parameter,
in addition to the bulk action contains a surface action in the form
$\int d^{D-1}x \sqrt{h} (c +\zeta K)\varphi ^2$, where the
integration goes over the brane, $h$ is the absolute value of the
determinant for the corresponding induced metric, $c$ is the brane
mass parameter for the field, and $K$ is the extrinsic curvature
scalar for the brane. This action gives $\delta $-type contributions
to the field equation located on the brane. Now the eigenfunctions
for the quantized bulk scalar field can be written in the form
(\ref{eigfunc1}), where the function $f(\xi )$ is a solution to the
equation which differs from (\ref{feq}) by the presence of the term
$-(c +\zeta K)\xi ^2 f(\xi )\delta (\xi -a)$ on the right hand side.
To obtain the boundary condition for the function $f(\xi )$ we
integrate the corresponding equation about $\xi =a$. Assuming that
the function $f(\xi )$ is continuous at this point one finds
\begin{equation}\label{discont}
    \lim _{\epsilon \to 0} \left. \frac{df}{d\xi }\right| _{\xi =
    a-\epsilon }^{\xi =a+\epsilon }=(c +\zeta K)f(a).
\end{equation}
For an untwisted scalar field we have $f(\xi )=f(2a-\xi )$ and from
(\ref{discont}) we obtain the boundary condition in the form
(\ref{bound1}) with
\begin{equation}\label{ABbraneworld}
   \frac{A}{B}=\frac{1}{2}\left( c-\frac{\zeta }{a}\right) ,
\end{equation}
where we have taken into account that for the boundary under
consideration $K=-1/a$. For a twisted scalar $f(\xi )=-f(2a-\xi )$
and from (\ref{discont}) we obtain the Dirichlet boundary condition.
Note that in the braneworld bulk the integration in the
normalization integral goes over two copies of the bulk manifold.
This leads to the additional coefficient $1/2$ in the expression
(\ref{Calfa}) for the normalization coefficient $C_{\alpha }$.
Hence, the Whightman function in the orbifolded braneworld case is
given by formula (\ref{W5}) with an additional factor $1/2$ in
formulae (\ref{W0}), (\ref{Wb1}). As it has been mentioned above
this function corresponds to the braneworld in the AdS black hole
bulk in the limit when the brane is close to the black hole horizon.

\section{Boundary-free geometry}

\label{sec:bfree}

In this section we will consider the vacuum expectation values for the
geometry without boundaries. First of all we note that the corresponding
Wightman function can be presented in the form%
\begin{eqnarray}
G_{0}^{+}(x,x^{\prime }) &=&\tilde{G}_{0}^{+}(x,x^{\prime })-\frac{%
r_{H}^{1-D}}{\pi ^{2}nS_{D}}\sum_{l=0}^{\infty }(2l+n)C_{l}^{n/2}(\cos
\theta )  \nonumber \\
&&\times \int_{0}^{\infty }d\omega e^{-\omega \pi }\cos [\omega (\tau -\tau
^{\prime })]K_{i\omega }(\lambda _{l}\xi )K_{i\omega }(\lambda _{l}\xi
^{\prime }),  \label{GM1}
\end{eqnarray}%
with the function%
\begin{eqnarray}
\tilde{G}_{0}^{+}(x,x^{\prime }) &=&\frac{r_{H}^{1-D}}{\pi ^{2}nS_{D}}%
\sum_{l=0}^{\infty }(2l+n)C_{l}^{n/2}(\cos \theta )  \nonumber \\
&&\times \int_{0}^{\infty }d\omega \cosh \{\omega \lbrack \pi -i(\tau -\tau
^{\prime })]\}K_{i\omega }(\lambda _{l}\xi )K_{i\omega }(\lambda _{l}\xi
^{\prime }).  \label{GM2}
\end{eqnarray}%
In this formula the $\omega $-integral can be evaluated with the result%
\begin{eqnarray}
\tilde{G}_{0}^{+}(x,x^{\prime }) &=&\frac{r_{H}^{1-D}}{2\pi nS_{D}}%
\sum_{l=0}^{\infty }(2l+n)C_{l}^{n/2}(\cos \theta )  \nonumber \\
&&\times K_{0}\left( \lambda _{l}\sqrt{\xi ^{2}+\xi ^{\prime 2}-2\xi \xi
^{\prime }\cosh (\tau -\tau ^{\prime })}\right) .  \label{GM3}
\end{eqnarray}%
It can be checked that this function is the Wightman function for the bulk
geometry $R^{2}\times S^{D-1}$ described by the line element%
\begin{equation}
ds^{2}=dt^{2}-(dx^{1})^{2}-r_{H}^{2}d\Sigma _{D-1}^{2},  \label{ds31}
\end{equation}%
where the coordinates $(t,x^{1})$ are related to the coordinates $(\tau ,\xi
)$ by formulas $t=\xi \sinh \tau $, $x^{1}=\xi \cosh \tau $. To see this we
note that the normalized eigenfunctions corresponding to this geometry are
given by the formula%
\begin{equation}
\widetilde{\varphi }_{\alpha }(x)=\frac{Y(m_{k};\vartheta ,\phi
)e^{ik_{1}x^{1}-i\omega _{l}t}}{\sqrt{4\pi \omega _{l}N(m_{k})r_{H}^{n+1}}},
\label{phialfM}
\end{equation}%
where $\alpha =(k_{1},m_{k})$ and $\omega _{l}^{2}=k_{1}^{2}+\lambda
_{l}^{2} $, with $\lambda _{l}$ defined by relation (\ref{lambdal}).
Substituting these functions into the corresponding mode sum and evaluating
the $k_{1}$-integral, for the case $|x^{1}-x^{1\prime }|>|t-t^{\prime }|$
one finds%
\begin{eqnarray}
\tilde{G}_{0}^{+}(x,x^{\prime }) &=&\sum_{\alpha }\widetilde{\varphi }%
_{\alpha }(x)\widetilde{\varphi }_{\alpha }^{\ast }(x^{\prime })  \nonumber
\\
&=&\frac{r_{H}^{1-D}}{2\pi nS_{D}}\sum_{l=0}^{\infty }(2l+n)C_{l}^{n/2}(\cos
\theta )K_{0}\left( \lambda _{l}\sqrt{(x^{1}-x^{1\prime })^{2}-(t-t^{\prime
})^{2}}\right) .  \label{GM4}
\end{eqnarray}%
Noting that $\xi ^{2}+\xi ^{\prime 2}-2\xi \xi ^{\prime }\cosh (\tau -\tau
^{\prime })=(x^{1}-x^{1\prime })^{2}-(t-t^{\prime })^{2}$\ we see that this
formula coincides with (\ref{GM3}).

In formula (\ref{GM1}), the divergences in the coincidence limit are
contained in the term $\tilde{G}_{0}^{+}(x,x^{\prime })$ and, hence, the
renormalization is needed for this term only. Now we turn to the evoluation
of the vacuum expectation values of the field square and the energy-momentum
tensor for the geometry $R^{2}\times S^{D-1}$ described by the line element (%
\ref{ds31}). The amplitude of the corresponding vacuum state we will denote
by $|\tilde{0}\rangle $. First of all note that from the problem symmetry it
follows that the expectation values $\langle \tilde{0}\vert \varphi
^{2}\vert \tilde{0}\rangle $, $\langle \tilde{0}t\vert T_{i}^{i}\vert \tilde{%
0}\rangle $ do not depend on the point of observation and%
\begin{eqnarray}
\langle \tilde{0}\vert T_{0}^{0}\vert \tilde{0}\rangle &=&\langle \tilde{0}%
\vert T_{1}^{1}\vert \tilde{0}\rangle ,  \label{vevEMTM} \\
\langle \tilde{0}\vert T_{2}^{2}\vert \tilde{0}\rangle &=&\cdots =\langle
\tilde{0}\vert T_{D}^{D}\vert \tilde{0}\rangle .  \nonumber
\end{eqnarray}%
The component $\langle \tilde{0}\vert T_{2}^{2}\vert \tilde{0}\rangle $ can
be expressed through the energy density by using the trace relation%
\begin{equation}
T_{i}^{i}=D(\zeta -\zeta _{c})\nabla _{i}\nabla ^{i}\varphi
^{2}+m^{2}\varphi ^{2}.  \label{TiiM}
\end{equation}%
From this relation it follows that%
\begin{equation}
\langle \tilde{0}\vert T_{2}^{2}\vert \tilde{0}\rangle =\frac{1}{D-1}\left(
m^{2}\langle \tilde{0}\vert \varphi ^{2}\vert \tilde{0}\rangle -2\langle
\tilde{0}\vert T_{0}^{0}\vert \tilde{0}\rangle \right) .  \label{vevT22M}
\end{equation}%
Hence, it is sufficient to find the renormalized vacuum expectation values
of the field square and the energy density. Using the eigenmodes (\ref%
{phialfM}), these quantities are presented as mode sums%
\begin{eqnarray}
\langle \tilde{0}\vert \varphi ^{2}\vert \tilde{0}\rangle &=&\frac{r_{H}^{-n}%
}{4\pi S_{D}}\int_{-\infty }^{+\infty }dk_{1}\sum_{l}\frac{D_{l}}{\eta
_{l}(r_{H}k_{1})},  \label{modephi2T00} \\
\langle \tilde{0}\vert T_{0}^{0}\vert \tilde{0}\rangle &=&\frac{r_{H}^{-n-2}%
}{4\pi S_{D}}\int_{-\infty }^{+\infty }dk_{1}\sum_{l}D_{l}\eta
_{l}(r_{H}k_{1}),  \label{modeT00}
\end{eqnarray}%
with the notation $\eta _{l}(x)=\sqrt{x^{2}+r_{H}^{2}\lambda _{l}^{2}}$. In
this formulas
\begin{equation}
D_{l}=(2l+D-2)\frac{\Gamma (l+D-2)}{\Gamma (D-1)l!}  \label{Dl}
\end{equation}%
is the degeneracy of each angular mode with given $l$. Of course, quantities
(\ref{modephi2T00}), (\ref{modeT00}) are divergent and some renormalization
procedure is needed. As such a procedure we will use the zeta function
technique. Let us define the zeta function%
\begin{eqnarray}
\zeta (s) &=&\int_{-\infty }^{+\infty }dx\sum_{l=0}^{\infty }D_{l}\eta
_{l}^{-2s}(x)  \nonumber \\
&=&\sqrt{\pi }\frac{\Gamma \left( s-\frac{1}{2}\right) }{\Gamma \left(
s\right) }\zeta _{S^{n+1}}\left( s-\frac{1}{2}\right) ,  \label{zetas}
\end{eqnarray}%
where
\begin{eqnarray}
\zeta _{S^{n+1}}(z) &=&\sum_{l=0}^{\infty }D_{l}(r_{H}\lambda _{l})^{-2z}
\nonumber \\
&=&\sum_{l=0}^{\infty }D_{l}\left[ (l+n/2)^{2}+b_{n}\right] ^{-z},
\label{zeta0s}
\end{eqnarray}%
is the zeta function for a scalar field on the spacetime $R\times S^{n+1}$
and
\begin{equation}
b_{n}=\zeta n(n+1)-n^{2}/4+m^{2}r_{H}^{2}.  \label{bn}
\end{equation}%
This function is well investigated in literature (see, for example, \cite%
{Camp90}) and can be presented as a series of incomplete zeta functions.
Here we recall that the function $\zeta _{S^{n+1}}(z)$ is a meromorphic
function with simple poles at $z=(n+1)/2-j$, where $j=0,1,2,\ldots $ for $n$
even and $0\leq j\leq (n-1)/2$ for $n$ odd. For $n$ even one has $\zeta
_{S^{n+1}}(-j)=0$, $j=1,2,\ldots $. Note that the function $\zeta
_{S^{n+1}}(z)$ can be expressed in terms of the function
\begin{equation}
F(z,c,b)=\sum_{l=1}^{\infty }\left[ (l+c)^{2}+b\right] ^{-z},  \label{Fzcb}
\end{equation}%
for $n$ even\ or its derivative for $n$ odd as%
\begin{equation}
\zeta _{S^{n+1}}(z)=\frac{2^{1-2\eta }}{\Gamma (D-1)}\sum_{j=0}^{[n/2]}\frac{%
b_{j}^{(\eta )}(b_{n})}{(j-z+2\eta )^{2\eta }}\left. \frac{\partial ^{2\eta }%
}{\partial c^{2\eta }}F(z-j-2\eta ,c,b_{n})\right\vert _{c=\eta },
\label{zeta0s1}
\end{equation}%
where $\eta =n/2-[n/2]$ and the square brackets mean the integer part of the
enclosed expression. In formula (\ref{zeta0s1}) the coefficients $%
b_{j}^{(\eta )}(b_{n})$ are defined by the relation%
\begin{equation}
\prod_{q=1}^{[n/2]}\left[ y-(q+\eta -1)^{2}-b_{n}\right] =%
\sum_{j=0}^{[n/2]}b_{j}^{(\eta )}(b_{n})y^{j}.  \label{bneta}
\end{equation}%
The formulas for the analytic continuation of the function $F(z,c,b)$ can be
found in \cite{Eliz94,Eliz90}.

Now on the base of formulas (\ref{modephi2T00}), (\ref{modeT00}) we have
\begin{equation}
\langle \tilde{0}\vert \varphi ^{2}\vert \tilde{0}\rangle =\frac{\zeta (1/2)%
}{4\pi S_{D}r_{H}^{D-1}},\quad \langle \tilde{0}\vert T_{0}^{0}\vert \tilde{%
0}\rangle =\frac{\zeta (-1/2)}{4\pi S_{D}r_{H}^{D+1}}.  \label{phi2T00}
\end{equation}%
By taking into account that the quantities $\zeta _{S^{n+1}}(0)$ and $\zeta
_{S^{n+1}}(-1)$ are finite, from formula (\ref{zetas}) we see that at $s=1/2$
and $s=-1/2$ the zeta function $\zeta (s)$ has simple poles with residues $%
\zeta _{S^{n+1}}(0)$ and $\zeta _{S^{n+1}}(-1)/2$, respectively. Hence, in
general, the vacuum expectation values of the field square and the energy
density contain the pole and finite contributions. The remained pole term is
a characteristic feature for the zeta function regularization method. Note
that for $n$ even $\zeta _{S^{n+1}}(-1)=0$ and the energy density is finite.

The vacuum expectation value of the field square in the boundary-free
geometry $Ri\times S^{D-1}$ is obtained from the Wightman function (\ref{GM1}%
) taking the coincidence limit. Using the relation%
\begin{equation}
C_{l}^{n/2}(1)=\frac{\Gamma (l+n)}{\Gamma (n)l!},  \label{Cl1}
\end{equation}%
for the corresponding quantity one finds%
\begin{equation}
\langle 0_{0}\vert \varphi ^{2}\vert 0_{0}\rangle =\langle
\tilde{0}\vert
\varphi ^{2}\vert \tilde{0}\rangle -\frac{r_{H}^{1-D}}{\pi ^{2}S_{D}}%
\sum_{l=0}^{\infty }D_{l}\int_{0}^{\infty }d\omega e^{-\omega \pi
}K_{i\omega }^{2}(\lambda _{l}\xi ),  \label{phi20}
\end{equation}%
where $|0_{0}\rangle $ is the amplitude for the corresponding vacuum state.
For large values of $\xi $, by using the asymptotic formulas for the
MacDonald function for large values of the argument, we can see that the
main contribution into the second term on the right of formula (\ref{phi20})
comes from the summand $l=0$ and we obtain%
\begin{equation}
\langle 0_{0}\vert \varphi ^{2}\vert 0_{0}\rangle =\langle \tilde{0}\vert
\varphi ^{2}\vert \tilde{0}\rangle -\frac{r_{H}^{1-D}e^{-2\lambda _{0}\xi }}{%
\pi ^{2}S_{D}\lambda _{0}\xi }.  \label{phi20a}
\end{equation}%
In the limit $\xi \rightarrow 0$ the second term on the right of (\ref{phi20}%
) diverges. For small values $\xi $ the main contribution comes from
large values $l$ and this term behaves as $(r_{H}/\xi )^{D-1}$.
Hence, near the horizon the boundary-free vacuum expectation value
of the field square is dominated by the second term on the right of
formula (\ref{phi20}) and is negative.

Now we turn to the vacuum expectation value of the energy-momentum tensor.
The corresponding operator we will take in the form%
\begin{equation}
T_{ik}=\partial _{i}\varphi \partial _{k}\varphi +\left[ \left( \zeta -\frac{%
1}{4}\right) g_{ik}\nabla _{l}\nabla ^{l}-\zeta \nabla _{i}\nabla _{k}-\zeta
R_{ik}\right] \varphi ^{2},  \label{EMT1}
\end{equation}%
with the trace relation (\ref{TiiM}). In (\ref{EMT1}) $R_{ik}$ is the Ricci
tensor for the bulk geometry and for the metric (\ref{ds22}) it has
components%
\begin{eqnarray}
R_{ik} &=&0,\quad i,k=0,1;\quad  \label{Rik} \\
R_{ik} &=&\frac{n}{r_{H}^{2}}g_{ik},\quad i,k=2,\ldots ,D.
\end{eqnarray}%
On the base of formula (\ref{EMT1}) the corresponding vacuum expectation
values are presented in the form%
\begin{equation}
\langle 0_{0}\vert T_{ik}\vert 0_{0}\rangle =\lim_{x^{\prime }\rightarrow
x}\nabla _{i}\nabla _{k}^{\prime }G_{0}^{+}(x,x^{\prime })+ \left[ \left(
\zeta -\frac{1}{4}\right) g_{ik}\nabla _{l}\nabla ^{l}-\zeta \nabla
_{i}\nabla _{k}-\zeta R_{ik}\right] \langle 0_{0}\vert \varphi ^{2}\vert
0_{0}\rangle .  \label{EMT2}
\end{equation}%
By using decomposition (\ref{GM1}), the vacuum energy-momentum tensor is
presented in the form%
\begin{equation}
\langle 0_{0}\vert T_{ik}\vert 0_{0}\rangle =\langle \tilde{0}\vert
T_{ik}\vert \tilde{0}\rangle +\langle T_{ik}\rangle ^{(0)},  \label{Tik0}
\end{equation}%
where the second summand on the right is given by formula (no summation over
$i$)%
\begin{equation}
\langle T_{i}^{k}\rangle ^{(0)}=-\frac{\delta
_{i}^{k}r_{H}^{1-D}}{\pi ^{2}S_{D}}\sum_{l=0}^{\infty
}D_{l}\lambda _{l}^{2}\int_{0}^{\infty }d\omega e^{-\omega \pi
}f^{(i)}\left[ K_{i\omega }(\lambda _{l}\xi )\right] .
\label{Tik00}
\end{equation}%
In this formula we use the notations%
\begin{eqnarray}
f^{(0)}[g(z)] &=&\left( \frac{1}{2}-2\zeta \right) \left[ \left( \frac{dg(z)%
}{dz}\right) ^{2}+\left( 1-\frac{\omega ^{2}}{z^{2}}\right) g^{2}(z)\right] +%
\frac{\zeta }{z}\frac{d}{dz}g^{2}(z)+\frac{\omega ^{2}}{z^{2}}g^{2}(z),
\label{f0} \\
f^{(1)}[g(z)] &=&-\frac{1}{2}\left( \frac{dg(z)}{dz}\right) ^{2}-\frac{\zeta
}{z}\frac{d}{dz}g^{2}(z)+\frac{1}{2}\left( 1-\frac{\omega ^{2}}{z^{2}}%
\right) g^{2}(z),  \label{f1} \\
f^{(i)}[g(z)] &=&\left( \frac{1}{2}-2\zeta \right) \left[ \left( \frac{dg(z)%
}{dz}\right) ^{2}+\left( 1-\frac{\omega ^{2}}{z^{2}}\right) g^{2}(z)\right] -%
\frac{\lambda _{l}^{2}-m^{2}}{(D-1)\lambda _{l}^{2}}g^{2}(z),  \label{fi}
\end{eqnarray}%
with $i=2,3,...,D$ and the indices 0 and 1 correspond to the coordinates $%
\tau $ and $\xi $. Note that for a minimally coupled scalar field the energy
density corresponding to (\ref{Tik00}) is negative for all values $\xi $. As
in the case of the field square, for large values $\xi $ the vacuum
expectation values (\ref{Tik00}) are exponentially suppressed by the factor $%
e^{-2\lambda _{0}\xi }$. For small values $\xi $ these expectation
values behave as $(r_{H}/\xi )^{D+1}$ and diverge on the horizon.
This type of horizon divergences is also found in the
well-investigated example of a bulk $Ri \times R^{D-1}$ (see
references \cite{Cand77}). Here the situation is similar to that
which takes place in the Schwarzshild black hole bulk when the
quantum field is prepared in the Boulware vacuum \cite{Cand80}. In
the black hole case quantities characterizing vacuum polarization
display this singular behavior because no physical realization is
possible of a static system whose size is arbitrarily close to the
gravitational radius. The nature of horizon divergences in
Rindler-like spacetimes is similar: no physical realization is
possible to a coordinate system for which the lines with a fixed
$\xi $ are arbitrarily close to the Rindler horizon. In this limit
the corresponding particles would move at infinitely large
acceleration.

\section{Vacuum expectation values induced by a spherical brane}

\label{sec:VEVEMT}

In this section we consider the vacuum expectation values induced by the
presence of a spherical brane. On the base of formula (\ref{W5}) for the
Whightman function, the vacuum expectation value\ for the field square is
presented in the form%
\begin{equation}
\langle 0|\varphi ^{2}|0\rangle =\langle 0_{0}|\varphi ^{2}|0_{0}\rangle
+\langle \varphi ^{2}\rangle ^{(b)},  \label{phi2b0}
\end{equation}%
where the second term on the right is induced by the spherical shell:
\begin{equation}
\langle \varphi ^{2}\rangle ^{(b)}=-\frac{r_{H}^{1-D}}{\pi S_{D}}%
\sum_{l=0}^{\infty }D_{l}\int_{0}^{\infty }d\omega \frac{\bar{K}_{\omega
}(\lambda _{l}a)}{\bar{I}_{\omega }(\lambda _{l}a)}I_{\omega }^{2}(\lambda
_{l}\xi ).  \label{phi2b}
\end{equation}%
This quantity is negative for Dirichlet boundary condition and is positive
for Neumann boundary condition. Similar formula can be derived for the
vacuum expectation value of the energy-momentum tensor. In accordance with
the relation (\ref{W5}) we can write%
\begin{equation}
\langle 0|T_{i}^{k}|0\rangle =\langle 0_{0}|T_{i}^{k}|0_{0}\rangle +\langle
T_{ik}\rangle ^{(b)},  \label{EMT3}
\end{equation}%
where $\langle 0_{0}|T_{i}^{k}|0_{0}\rangle $ is the vacuum expectation
value for the situation without the spherical shell and $\langle
T_{ik}\rangle ^{(b)}$ is induced by the presence of the sphere. The latter
is finite for the points away from the sphere surface and the horizon, and
is given by formula (no summation over $i$)%
\begin{equation}
\langle T_{i}^{k}\rangle ^{(b)}=-\frac{\delta _{i}^{k}r_{H}^{1-D}}{\pi S_{D}}%
\sum_{l=0}^{\infty }D_{l}\lambda _{l}^{2}\int_{0}^{\infty }d\omega \frac{%
\bar{K}_{\omega }(\lambda _{l}a)}{\bar{I}_{\omega }(\lambda _{l}a)}F^{(i)}%
\left[ I_{\omega }(\lambda _{l}\xi )\right] ,  \label{EMT4}
\end{equation}%
where the expressions for the functions $F^{(i)}[g(z)]$ are obtained from
the corresponding formulas for the functions $f^{(i)}[g(z)]$ replacing $%
\omega \rightarrow i\omega $. As it has been explained in section
\ref{sec:WF}, the corresponding quantities in the orbifolded
braneworld version of the problem are obtained from (\ref{phi2b}),
(\ref{EMT4}) with an additional coefficient $1/2$ and boundary
coefficients (\ref{ABbraneworld}).

Now let us consider various limiting cases of the general formulas for the
brane-induced vacuum expectation values. In the limit $\xi \rightarrow a$
the expectation values for both field square and the energy-momentum tensor
diverge. These surface divergences are well known in quantum field theory
with boundaries and are investigated for various type of boundary conditions
and geometries. For the points near the brane the main contributions come
from large values $l$. Using the uniform asymptotic expansions for the
Bessel modified function, to the leading order one finds%
\begin{equation}
\langle \varphi ^{2}\rangle ^{(b)}\approx -\frac{\delta _{B}\Gamma \left(
\frac{D-1}{2}\right) }{(4\pi )^{\frac{D+1}{2}}(a-\xi )^{D-1}},
\label{phi2close}
\end{equation}%
for the field square and%
\begin{equation}
\langle T_{0}^{0}\rangle ^{(b)}\approx \langle T_{2}^{2}\rangle
^{(b)}\approx \frac{D(\zeta -\zeta _{c})\delta _{B}}{2^{D}\pi ^{\frac{D+1}{2}%
}(a-\xi )^{D+1}}\Gamma \left( \frac{D+1}{2}\right) ,  \label{T00close}
\end{equation}%
for the components of the energy-momentum tensor, $\zeta _{c}=(D-1)/4D$ is
the curvature coupling parameter for a conformally coupled scalar, and we
have introduced the notation
\begin{equation}
\delta _{B}=\left\{
\begin{array}{cc}
1, & B=0 \\
-1, & B\neq 0%
\end{array}%
\right. .  \label{deltaB}
\end{equation}%
These leading terms are the same as those for a flat brane in the Minkowski
bulk. They do not depend on the mass and Robin coefficients and have
opposite signs for Dirichlet and non-Dirichlet boundary conditions. The
leading term in the asymptotic expansion of the component $\langle
T_{1}^{1}\rangle ^{(b)}$ is obtained from (\ref{T00close}) by using
covariant continuity equation for the tensor $\langle T_{i}^{k}\rangle
^{(b)} $. This term behaves as $(a-\xi )^{-D}$.

For large values of the ratio $a/r_{H}$ the quantity $\lambda _{l}a$ is
large and we can replace the Bessel modified functions with this argument by
their asymptotics for large values of the argument. This leads to the
formulas (no summation over $i$)%
\begin{eqnarray}
\langle \varphi ^{2}\rangle ^{(b)} &\approx &-\frac{e^{-2\lambda _{0}a}}{%
S_{D}r_{H}^{D-1}}\frac{A-B\lambda _{0}}{A+B\lambda _{0}}\int_{0}^{\infty
}d\omega \,I_{\omega }^{2}(\lambda _{0}\xi ),  \label{phi2far} \\
\langle T_{i}^{k}\rangle ^{(b)} &\approx &-\frac{\delta _{i}^{k}\lambda
_{0}^{2}e^{-2\lambda _{0}a}}{S_{D}r_{H}^{D-1}}\frac{A-B\lambda _{0}}{%
A+B\lambda _{0}}\int_{0}^{\infty }d\omega \,F^{(i)}[I_{\omega }(\lambda
_{0}\xi )],  \label{Tikfar}
\end{eqnarray}%
with the exponential suppression of the brane-induced vacuum expectation
values.

In the near horizon limit, $\xi /r_{H}\ll 1$, with fixed $a/r_{H}$, the main
contributions into the $\omega $-integrals come from small values $\omega $.
Expanding the functions $I_{\omega }^{2}(\lambda _{l}\xi )$, to the leading
order one finds (no summation over $i$)%
\begin{eqnarray}
\langle \varphi ^{2}\rangle ^{(b)} &\approx &-\frac{r_{H}^{1-D}{\cal I}(a)}{%
2\pi S_{D}\ln (2r_{H}/\xi )},\quad {\cal I}(a)=\sum_{l=0}^{\infty }D_{l}%
\frac{\bar{K}_{0}(\lambda _{l}a)}{\bar{I}_{0}(\lambda _{l}a)}
\label{phi2bnearhor} \\
\langle T_{0}^{0}\rangle ^{(b)} &\approx &-\langle T_{1}^{1}\rangle
^{(b)}\approx -\frac{\zeta r_{H}^{1-D}{\cal I}(a)}{2\pi S_{D}\xi ^{2}\ln
^{2}(2r_{H}/\xi )},  \label{T00nearhor} \\
\langle T_{i}^{i}\rangle ^{(b)} &\approx &\frac{(4\zeta -1)r_{H}^{1-D}{\cal I%
}(a)}{4\pi S_{D}\xi ^{2}\ln ^{3}(2r_{H}/\xi )},\quad i=2,3,\ldots .
\label{Tiinearhor}
\end{eqnarray}%
As we see the brane-induced part in the vacuum expectation value of the
field square vanishes at the horizon, whereas the expectation values of the
energy-momentum tensor diverge. Recall that near the horizon the boundary
free part of the energy-momentum tensor behaves as $\xi ^{-D-1}$ and the
vacuum expectation values are dominated by this part. Note that the function
${\cal I}(a)$ is positive for Dirichlet boundary condition and is negative
for Neumann boundary condition. In the large mass limit the brane induced
vacuum expectation values are exponentially suppressed by the factor $%
e^{-2m(a-\xi )}$.

\section{Conclusion}

\label{sec:Conc}

In this paper, we investigate the quantum vacuum effects produced by a
spherical brane in the $(D+1)$-dimensional bulk $Ri\times S^{D-1}$, with $Ri$
being a two-dimensional Rindler spacetime. The corresponding line element (%
\ref{ds22}) describes the near horizon geometry of a non-extremal black hole
spacetime defined by the line element (\ref{ds21}). The case of a massive
scalar field with general curvature coupling parameter and satisfying the
Robin boundary condition on the sphere is considered. To derive formulas for
the vacuum expectation values of the square of the field operator and the
energy-momentum tensor, we first construct the positive frequency Wightman
function. This function is also important in considerations of the response
of a particle detector at a given state of motion through the vacuum under
consideration \cite{Birr82}. The Wightman function is presented as the sum
of the Whightman function for the boundary-free geometry and the term
induced by the presence of the spherical brane. For the points away the
boundary and horizon the divergences in the coincidence limit are contained
in the first term and hence, the renormalization is needed for this term
only. In section \ref{sec:bfree} we have shown that the the Whightman
function for the boundary-free $Ri\times S^{D-1}$ geometry can be presented
in the form of a sum of the Whightman function for the boundary-free $%
R^{2}\times S^{D-1}$ geometry plus a term which is finite in the coincidence
limit. As a result for the renormalization of the vacuum expectation values
of the field square and the energy-momentum tensor it is sufficient to
renormalize the corresponding quantities for the geometry $R^{2}\times
S^{D-1}$. The latter are point-independent and as a renormalization
procedure we employ the zeta function regularization method. The
corresponding zeta function can be expressed in terms of the zeta function $%
\zeta _{S^{n+1}}(z)$ in the geometry $R\times S^{n+1}$, the analytic
continuation for which is well investigated in literature. Alternatively the
zeta function can be expressed in terms of the function (\ref{Fzcb}). The
vacuum expectation values of the field square and the energy-momentum tensor
are expressed in terms of the zeta function by formulas (\ref{phi2T00}). In
general, they contain pole and finite contributions. In the minimal
subtraction scheme the pole terms are omitted. As a result the vacuum
expectation values of the field square and the energy-momentum tensor for
the boundary-free $Ri\times S^{D-1}$ geometry are determined by formulas (%
\ref{phi20}), (\ref{Tik0}), (\ref{Tik00}). On the horizon these expectation
values diverge. The leading terms in the near horizon asymptotic expansions
behave as $(r_{H}/\xi )^{D-1}$ for the field square and as $(r_{H}/\xi
)^{D+1}$ for the components of the energy-momentum tensor. The vacuum
expectation values induced by a spherical brane in the bulk geometry $%
Ri\times S^{D-1}$ are investigated in section \ref{sec:VEVEMT}. Near the
brane the vacuum expectation values are dominated by the boundary parts and
the corresponding components diverge at the brane. For non-conformally
coupled scalars the leading terms in the corresponding asymptotic expansions
are given by formulas (\ref{phi2close}), (\ref{T00close}) and are the same
as those for an infinite plane boundary in the Minkowski bulk. These terms
do not depend on the mass and Robin coefficients and have opposite signs for
Dirichlet and non-Dirichlet boundary conditions. For large values of the
ratio $a/r_{H}$ the brane-induced vacuum expectation values are
exponentially suppressed by the factor $\exp [-2(a/r_{H})\sqrt{\zeta
n(n+1)+m^{2}r_{H}^{2}}]$. For the points near the horizon one has $\xi
/r_{H}\ll 1$ and the brane-induced vacuum expectation value of the field
square vanishes as $\ln ^{-1}(2r_{H}/\xi )$. Unlike to the field square, the
brane-induced parts in the vacuum expectation values of the energy-momentum
tensor diverge on the horizon. In this paper we have considered the vacuum
expectation values induced by a spherical brane in the region $0<\xi <a$. By
the similar way the corresponding quantities for the region $\xi >a$ may be
investigated. It can be seen that in this region the brane-induced
quantities can be obtained from those for the region $0<\xi <a$ by the
replacements $I_{\omega }\leftrightarrows K_{\omega }$ in formulas (\ref%
{phi2b}) and (\ref{EMT4}). In the corresponding braneworld scenario
the geometry is made up by two slices of the region $0<\xi <a$ glued
together at the brane with a orbifold-type symmetry condition and
the ratio $A/B$ for bulk scalars is related to the brane mass
parameter of the field by formula (\ref{ABbraneworld}). The
corresponding formulae for the Wightman function and the vacuum
expectation values of the field square and the energy-momentum
tensor are obtained from those given above with an additional
coefficient $1/2$. They describe the braneworld in the AdS black
hole bulk in the limit when the brane is close to the black hole
horizon.

Note that in this paper we have considered quantum vacuum effects
induced by boundaries in a prescribed background, i.e. the
gravitational back-reaction of quantum effects is not taken into
account. This back-reaction could have a profound effect on the
dynamical evolution of the bulk model. We do not consider this
important extension of the theory, but note that the results
presented here constitute the starting point for such
investigations.

\section*{Acknowledgement}

The work of AAS was supported by the Armenian National Science and Education
Fund (ANSEF) Grant No. 05-PS-hepth-89-70 and in part by the Armenian
Ministry of Education and Science Grant No. 0124. AAS is grateful to the
Fund Robert Boghossian \& Sons.

\end{document}